\documentstyle[twocolumn,prl,aps,graphics,tighten,floats]{revtex} 

\newcommand{\ruth}{Sr$_2$RuO$_4$} 
\newcommand{\tc}{$T_c$} 
\newcommand{\oo}{$\langle$110$\rangle$} 
\newcommand{\on}{$\langle$100$\rangle$}
\newcommand{\nmodla}{\bar\eta_{\text{L100}}} 
\newcommand{\nmodlb}{\bar\eta_{\text{L110}}} 
\newcommand{\nmodta}{\bar\eta_{\text{T100}}} 
\newcommand{\nmodtb}{\bar\eta_{\text{T110}}} 
\newcommand{\nmodm}{\bar\eta_{\text{mode}}}
\newcommand{\etal}{{\it et al.}}
\newcommand{\angdeg}{$^{\circ}$}

\begin{document} 

\wideabs{ 
 \draft 
 \preprint{Submitted to PRL} 
 \title{Ultrasound Attenuation in Sr$_2$RuO$_4$: an Angle-Resolved
        Study of the Superconducting Gap Function}
 \author{C. Lupien, W.A. MacFarlane, Cyril Proust and Louis
    Taillefer }
 \address{Canadian Institute for Advanced Research\\
    Department of Physics, University of Toronto, Toronto, Canada M5S
    1A7} 
 \author{Z.Q. Mao and Y. Maeno} 
 \address{Department of Physics, Graduate School of Science,
   Kyoto University, Kyoto 606-8502, Japan\\
   CREST, Japan Science and Technology Corporation} 
 \date{April 4, 2001}
 \maketitle

 \begin{abstract}
   We present a study of the electronic ultrasound attenuation $\alpha$ in the
   unconventional superconductor \ruth .  The power law behavior of
   $\alpha$ at temperatures down to $T_c/30$ clearly indicates
   the presence of nodes in the gap.
   In the normal state, we find an enormous anisotropy of
   $\alpha$ in the basal plane of the tetragonal structure.
   In the superconducting state, the temperature dependence of $\alpha$
   also exhibits significant anisotropy. We discuss these results in
   relation to possible gap functions.
 \end{abstract} 
}

The symmetry of the order parameter $\Delta$ is of fundamental importance in
understanding the nature of superconductivity. To
this day, only two symmetries have been unambiguously identified
amongst all known superconductors: $s$-wave symmetry in conventional
superconductors and $d$-wave symmetry in some of the high-\tc\ 
cuprates.  Both cases involve spin-singlet Cooper pairing.
Heavy-fermion and
organic superconductors are widely believed to be
unconventional (i.e.\  not $s$-wave); however, for
a variety of technical reasons, the symmetry
of $\Delta$ has not been firmly established in
any of these systems.
In this context, \ruth\  has emerged as a highly
promising candidate for the study of unconventional superconductivity
\cite{Maeno94} because of its 
thoroughly characterized  Fermi Surface (FS)
\cite{fermi} and the
availability of sizable high-quality crystals.
Moreover, it may well prove to be the
first clear example of a spin-triplet superconductor \cite{Rice95}.

The most direct evidence for spin-triplet pairing is the absence of
detectable change in
the Pauli susceptibility below \tc\
as measured by $^{17}$O Knight shift \cite{Ishida98} and 
neutron scattering \cite{Duffy00}.
In addition, the appearance of a small random spontaneous
internal field in the superconducting state \cite{Luke98}
indicates that $\Delta$ breaks time-reversal symmetry,
a conclusion supported by analysis of the square
flux lattice \cite{Kealey00}.
The simplest $\Delta$ consistent with
these two properties possesses
an orbital wave function ${\bf d}={\bf \hat{z}}(k_x + i k_y)$.
This $p$-wave state has an isotropic gap (without nodes),
and for several years
it was viewed as the leading candidate for the superconducting
order parameter of
\ruth\ \cite{Maeno01}.

However, in the last year it has become clear that an
isotropic gap is inconsistent with the power law $T$
dependence of several quantities observed deep in the superconducting
state \cite{Nishizaki00,kappa,Ishida00,Bonalde00}.
Such behavior is most naturally explained by line nodes in the gap.
Thus the detailed symmetry of the order parameter in \ruth\ is still 
very much an
open question. In order to determine the {\it orbital} symmetry of $\Delta$,
one requires an angle-resolved probe to reveal the location of
nodes in the gap on the FS.
One approach is to seek an anisotropic response to an applied magnetic
field.
Such an effect has been observed in heat transport with the field rotating in
the basal plane. However, the anisotropy was quite small and was deemed
incompatible with vertical line nodes \cite{kappa}. In the study presented in
this Letter we use the intrinsic angular resolution of ultrasound attenuation
to directly probe the anisotropy of the quasiparticle excitation spectrum of
superconducting \ruth\  in zero magnetic field.

Moreno and Coleman have emphasized the power of ultrasound attenuation
in studies of anisotropic superconductors \cite{Moreno96}.
In their model calculations, performed in the hydrodynamic limit where
the electron mean free path $l$ is much shorter than the sound
wavelength $\lambda$, the low $T$ power law behavior of $\alpha(T)$ is
shown to depend strongly on the direction of sound propagation ${\bf \hat
q}$ and polarization ${\bf \hat e}$ relative to the nodes. For example, in
the case of a 2D gap with $k_x^2-k_y^2$ symmetry and for transverse
sound with ${\bf \hat q}$ and ${\bf \hat e}$ in the 
basal plane, quasiparticles at
nodes along the \oo\  are ``inactive'' when ${\bf \hat q} \parallel $ \oo ,
yielding a strong $T^{3.5}$ power law, whereas they are maximally
``active'' in attenuating sound for ${\bf \hat q} \parallel $ \on , 
resulting in
a weak $T^{1.5}$ dependence.  In this way ultrasound attenuation
experiments can locate nodes in the gap, as was done for
the heavy-fermion superconductor UPt$_3$ \cite{Ellman96}.

We have performed measurements of longitudinal and transverse
ultrasound attenuation in \ruth\  in the temperature range 0.04~--~4~K,
with sound propagating along the \on\  and \oo\  directions in the basal
plane of the tetragonal crystal structure.
Our experiments were carried out on oriented pieces cut from a single
high-quality crystal grown by the travelling
solvent float zone
technique \cite{samp}. \tc\  defined as the peak of the magnetic
susceptibility ($\chi''$) is 1.37~K.
Samples were polished to 1 $\mu$m roughness, with two
opposite faces whose parallelism was estimated to be better than 1.5
$\mu$m/mm. The alignment of the polished faces relative to the crystal
axes was determined by Laue back reflection to be less than
0.5\angdeg\   off axis.  (For propagation along \on , we estimate the
misalignment to be 0.2--0.3\angdeg .)  The crystals were 3.98 and
4.23 mm long, and had overlapping regions of the two parallel faces of
$\sim$~2 mm$^2$.  Pulse-echo ultrasound was performed using a
homebuilt spectrometer and commercial (30 MHz) LiNbO$_3$ transducers,
bonded to the crystals with a thin layer of grease.  With this technique, the
absolute sound velocities can be measured to a few percent, but only a
relative measurement of the electronic attenuation is possible because
one cannot distinguish power losses from other origins.
We denote the modes as follows: T100, T110, L100, and L110, where the
polarization (T for transverse, L for longitudinal) is always in the
plane. The indices denote the propagation direction.  From the echo
spacing, the sound velocities were found to be 3.30, 2.94,
6.28, and 6.41 mm/$\mu$s, respectively, within 3\%.  (These are
somewhat different from previous reports \cite{Matsui}.) 
Typically at low temperature for both the
T100 and L110 modes we recorded more than 50 echoes, while L100 had
at least 20 and T110 at least 2. Preliminary measurements were
accomplished in a $^3$He cryostat and were consistent with data taken in
a dilution refrigerator up to 4 K. The results were also 
independent of excitation power. 

\begin{figure}[tbh]
\resizebox{\linewidth}{!}{\includegraphics{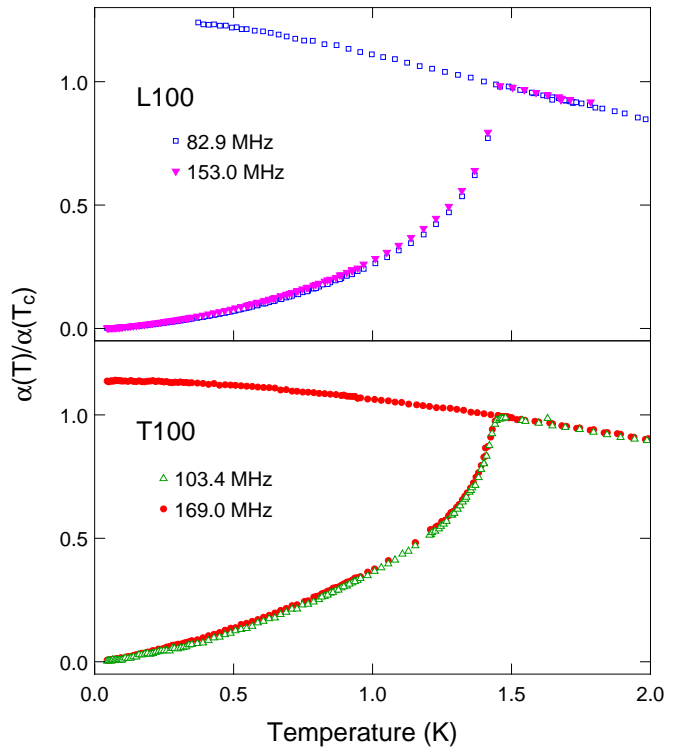}}
\vskip 0.1cm  
\caption[]{ 
  Ultrasound attenuation $\alpha(T)$ normalized at \tc , both in the
  superconducting state ($B = 0$) and in the normal state ($B =
  1.5~{\text{T}} \perp c$). We show both T100 and L100 at different
  frequencies to demonstrate that the shape of the curves does not
  depend on frequency.  }
\label{freqfig}
\end{figure}

An important parameter that controls the behavior of the attenuation
is the ratio $\lambda/l$.
Theoretical approaches are quite different
depending on the value of this parameter.
The electronic mean free
path below \tc\  is on the order of 1
$\mu$m, as estimated from either \tc\  or resistivity
\cite{Mao99}.  From this 
we expect that the crossover region $ql \sim 1$ (where
$q=2\pi /\lambda$ is the sound wavevector) from the hydrodynamic to
the quantum regime occurs at 0.5 (1.0) GHz for the transverse
(longitudinal) modes.  The theoretical dependence of $\alpha$ on frequency is
$\nu^2$ in the hydrodynamic regime and $\nu$ in the quantum regime.
Fig.~\ref{freqfig} shows data for two modes at two frequencies,
normalized at \tc .  The values of $\alpha(T_c)$ obey approximate
$\nu^2$ scaling, and the shape of $\alpha(T)$ is independent of $\nu$
and reproducible at many frequencies up to 500 MHz.
We conclude that all our data are in the hydrodynamic limit.

\begin{figure}[tbh]
\resizebox{.99\linewidth}{!}{\includegraphics{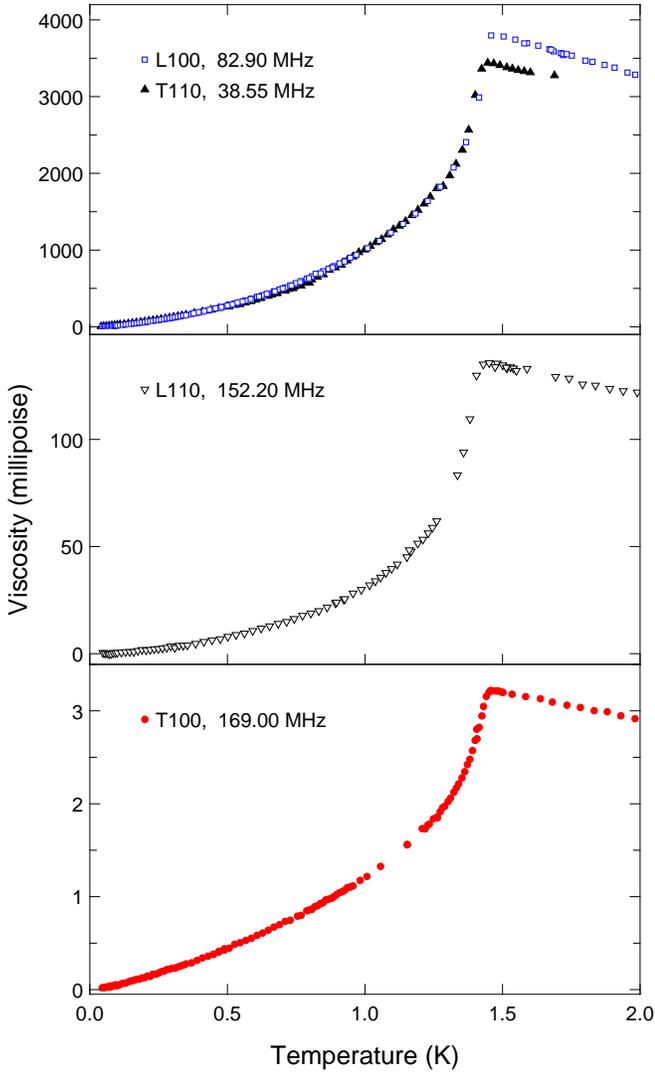}}
\vskip 0.1cm  
\caption[]{ Viscosity for the four in-plane modes. The viscosity $\eta$
  is related to the attenuation by $\eta=\alpha\rho
  c_s^3/(2\pi\nu)^2$, using a density $\rho=5.9$ g/cm$^3$ and the
  sound velocities ($c_s$) given in the text (poise $\equiv$ g/cm s).
  The data is extrapolated to 0~K to obtain the zero of the vertical
  axis.}
\label{rawfig} 
\end{figure}

In this limit, the attenuation is related to the 4$^{th}$ rank
electronic viscosity tensor ($\eta$) via the relation \cite{rodr85}:
\begin{equation} 
\alpha =\frac{(2\pi \nu)^2}{\rho c_s^3} \eta_{ijkl}
         \hat e_i \hat q_j \hat e_k \hat q_l = 
  \frac{(2\pi \nu)^2}{\rho c_s^3} \bar\eta_{\text{mode}}, 
\end{equation}
where $\rho$ is the mass density and $c_s$ is the sound velocity for
the mode with polarization and propagation directions ${\bf \hat e}$ and
${\bf \hat q}$. This defines $\nmodm$ as the linear combination of
$\eta_{ijkl}$ components for a particular mode. Under tetragonal
symmetry there are 6 independent components of $\eta$, but only three
determine the attenuation of the four in-plane modes, so that
$\nmodm$ is given by
 $\eta_{1212}$, $\eta_{1111}$, 
 $\frac{1}{2}(\eta_{1111}-\eta_{1122})$, and
 $\frac{1}{2} (\eta_{1111}+\eta_{1122}+2\eta_{1212})$ 
for T100, L100, T110, and L110, respectively.
We present $\nmodm$ in Fig.~\ref{rawfig}.
The curves show that L100 and T110 have the same
shape and almost the same amplitude.
The L110 mode has a similar
shape but a smaller amplitude.  However, the $T$ dependence of the
T100 mode is clearly weaker and the magnitude of $\nmodta$ is 
{\it about 1000 times smaller} than $\nmodla$ or $\nmodtb$.
We note that the alignment of the crystal faces mentioned above is sufficient
to yield the intrinsic $\nmodta$ \cite{align-note}. Such a dramatic anisotropy
of the normal state electronic $\eta$ is unprecedented. 
Next we consider possible origins for this anisotropy.

The viscosity is related to the deformation of the
FS caused by the sound wave's strain field, weighted by the
electronic relaxation time $\tau$ \cite{mason58}. 
The anisotropy of
$\eta$ may in part come from angular dependence of $\tau$ or 
the topology of the FS which
is composed of three separate sheets 
(labelled $\alpha$, $\beta$, and $\gamma$) \cite{fermi}.
Although the $\gamma$ sheet is fairly isotropic, the $\alpha$ and
$\beta$ sheets have flat regions that could make significantly
different contributions to the different components of $\eta$.
Interband effects at the point of near contact between the three
sheets may also be important. 
However, the anisotropy is so large that
it probably requires significant variation in the electron-acoustic phonon
coupling for the different modes,
possibly originating in the highly directional nature of the Ru $4d$ / O $2p$
hybrid orbitals that compose the conduction bands.

To facilitate discussion of $\alpha(T)$ below \tc ,
we continue with some remarks about the observed viscosities.
Theory of 2D electronic viscosity \cite{Graf00-2} suggests a close
relation between two of the relevant components, namely
$\eta_{1111}=-\eta_{1122}$ for certain simple FS.
The data presented in the top panel of Fig.~\ref{rawfig} shows
that this relation is nearly fulfilled in \ruth , 
in both magnitude {\it and} $T$ dependence.
The slight imperfection of this symmetry should not be disturbing, 
since, e.g.,
the real FS is not exactly 2D.
This relation also implies that $\nmodm$ in the bottom two
panels should be nearly equal. The apparent dramatic violation of this
prediction is, however, a trivial consequence of the slight imperfection of
the symmetry together with the enormous anisotropy ($|\eta_{1212}| \ll
|\eta_{1111}|,|\eta_{1122}|$).
While $\nmodlb$ contains all three viscosities, it is dominated by
the slight inequality between $\nmodla$ and $\nmodtb$ (as the
difference is still much larger than $\eta_{1212}$). Because the
observed $\nmodlb (T)$ is so similar to $\nmodla (T)$, this
difference is primarily one of magnitude.
Thus there are effectively only two $T$ dependences for the
in-plane viscosities, one represented exclusively by $\nmodta$ and the
other most clearly by $\nmodla$. We compare these in detail next.

\begin{figure}[tbh]
\resizebox{\linewidth}{!}{\includegraphics{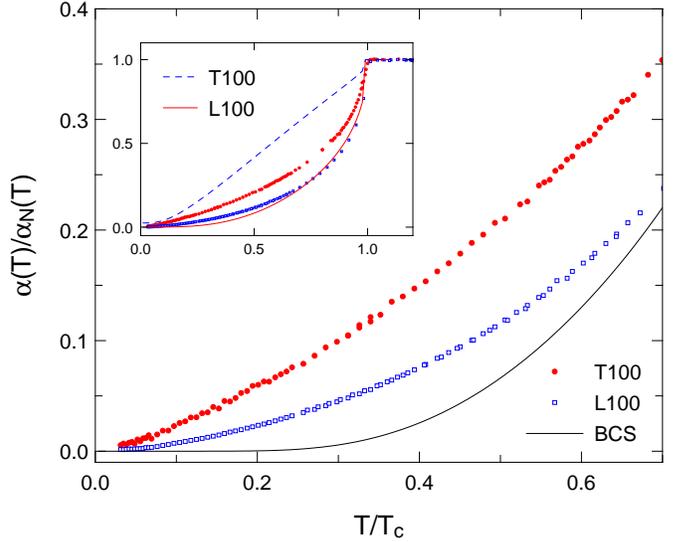}}
\vskip 0.1cm   
\caption[]{   
  $\alpha(T)$ divided by the normal state attenuation
   {\it vs} reduced temperature. The solid line
  is the simple BCS prediction.  {\it Inset}: data up to \tc ,
  compared with calculations for $f$-wave discussed in the text
  \protect\cite{Graf00-2}.
}
\label{lowtfig}    
\end{figure} 

The attenuation for these two modes is shown in Fig.~\ref{lowtfig},
normalized by their normal state curves $\alpha_N(T)$
measured in a magnetic field (1.5~T, approximately in-plane)
exceeding the upper critical field.
A simple BCS model (weak-coupling isotropic
$s$-wave gap) or a clean fully gapped $p$-wave state  \cite{Kee00}
give an exponential decrease of $\alpha(T)$ at low $T$,
such that below $\sim T_c/4$ the attenuation is essentially zero
(solid line in Fig.~\ref{lowtfig}).
This is obviously not the case here, where we see power
laws persisting down to $T_c/30$, {\it providing compelling evidence
for the presence of nodes in the gap}.  It is also clear that there is
anisotropy in the $T$ dependence of the viscosity in the basal plane. The 
weakest $T$ dependence occurs for the T100 mode.
Below 0.7~K, the normalized $\nmodta$ and $\nmodla$ are well described by
the power laws 0.53$t^{1.4}$ and 0.38$t^{1.8}$ respectively, where $t=T/T_c$.
The latter power law is consistent with another recent measurement of
$\nmodtb$ \cite{matsui01}.
The anisotropy revealed in Fig.~\ref{lowtfig} is
much larger than any other kind of anisotropy observed so far in the basal
plane of superconducting \ruth .
Next we discuss the anisotropy of $\alpha(T)$ in relation
to the nodal structure of the gap.

At the outset we note that the huge anisotropy in the normal state viscosity
is not included in theories forming the basis of existing
calculations of $\alpha(T)$ in the superconducting state. Thus, we can only
draw tentative conclusions from a comparison of these theories to the
data.
We identify two types of order parameter that have been proposed:
those with vertical line nodes (e.g.\  \cite{Graf00-2}) and
those with horizontal line nodes (e.g.\  \cite{Hasegawa00,Zhit01}).
We now consider these in turn.

If the angular sensitivity articulated by Moreno and Coleman survives
in a theory correctly accounting for the normal state viscosity,
then the anisotropy of
Fig.~\ref{lowtfig} is naturally interpreted in terms of vertical nodal lines.
In this case, the observed anisotropy of the
$T$ dependence between L100 and T100 is qualitatively consistent with a
$\Delta$ of $k_x^2-k_y^2$ orbital symmetry in 2D such as 
found in the cuprates.
However, given that a $d$-wave order parameter is inconsistent 
with the combination of
spin-triplet symmetry and time-reversal symmetry breaking observed in \ruth ,
one needs to invoke {\it accidental} nodes, such as those 
of an $f$-wave state with
${\bf d} = {\bf \hat{z}}(k_x + i k_y)(k_x^2-k_y^2)$ \cite{Hasegawa00}.
In the inset of Fig.~\ref{lowtfig}, we compare
our normalized data for L100 and T100 with the calculation of Graf and
Balatsky \cite{Graf00-2} for such an $f$-wave gap
on a single isotropic 2D FS \cite{dxy-note}.
In detail the $T$ dependences differ significantly from the
simple model predictions. 
There is good overall agreement for the L100 mode;
however, the low $T$ power law exponent
is much lower than expected (1.8 vs 3.5).
The data for the T100 mode is
always below the calculated curve,
but, in contrast, the exponent is close to the
predicted 1.5 \cite{Moreno96,Graf00-2}.
These significant differences may be the consequence of the real FS
of \ruth\  or of a more complex nodal structure. 
Detailed calculations are needed to evaluate how strongly the observed
anisotropic $T$ dependence favors such a $\Delta$.

For a simple horizontal line node,
one would not expect anisotropy in the plane. Thus we must presume a
separate source of the observed anisotropy.
One possibility is a (non-nodal) 4-fold variation of $|\Delta|$ in the plane.
The magnitude of the modulation required to explain our data
could be computed and the resulting gap function compared with other
measurements such as the electronic specific heat.
However, at sufficiently low $T$, we still expect the power law
exponent to be isotropic in this scenario.

In summary, the power law $T$ dependence of the ultrasound attenuation
persisting down to $T_c/30$ establishes unambiguously the existence of
nodes in the gap function of \ruth .  These power laws and the
anisotropy of the attenuation in the superconducting state
excludes the possibility of a pure $p$-wave state,
but may prove to be consistent with a state that has either vertical nodes
on the diagonal or a horizontal line of nodes in conjunction with
significant 4-fold gap modulation.
A detailed calculation taking into
account the real (multi-sheet) FS of \ruth\  and the large
normal state anisotropy is needed to
properly extract the full order-parameter information contained in the
anisotropy and temperature dependence of $\alpha(T)$.
 
We would like to thank M. J. Graf for useful discussions and R. W.  Hill
for his help with the experiments. This work was supported by the
Canadian Institute for Advanced Research and funded by NSERC.
CL acknowledges the support of a Walter C. Sumner
Fellowship and a FCAR scholarship (Qu\'ebec), and LT the support of
a Premier's Research Excellence Award (Ontario).




\begin{thebibliography}{99}
\bibitem{align-note} The other components mix
in to $\nmodm$ as $\sin^2(2\theta)$, where $\theta$ is the small in-plane
misalignment angle.
\end{thebibliography}
\end{document}